\documentclass{fairmeta}
\usepackage{amsmath}
\usepackage{enumerate} 
\usepackage{bm}
\usepackage{algorithm}
\usepackage{floatflt}
\usepackage{algpseudocode}
\usepackage{amsfonts}
\usepackage{amsthm}
\usepackage{newtxtt}
\usepackage{algorithm}
\usepackage{colortbl} 
\usepackage{cleveref}
\usepackage{diagbox} 
\usepackage[utf8]{inputenc}
\usepackage{textgreek}
\usepackage{colortbl}
\usepackage{nicematrix}
\usepackage{makecell}
\usepackage{float}
\usepackage{arydshln}
\usepackage[frozencache,cachedir=.]{minted}
\usepackage{caption}
\usepackage{subcaption}
\usepackage{tcolorbox}
\usepackage{amssymb}
\usepackage{xspace}
\usepackage{wrapfig}
\usepackage{adjustbox}
\usepackage{tabularx}
\usepackage{booktabs}
\usepackage{mathtools}
\usepackage{wrapfig}
\usepackage{amssymb}
\usepackage{graphicx}

\usepackage{silence}
\makeatletter
\patchcmd{\wrong@fontshape}{\@gobbletwo}{}{}{}
\makeatother
\definecolor{upColor}{RGB}{17,138,21}
\definecolor{downColor}{RGB}{174,36,67}

\newtheorem{theorem}{Theorem}[]

\newtheorem{remark1}[theorem]{Remark}

\title{Generative Video Compression: Towards 0.01\% Compression Rate for Video Transmission}

\author[]{Xiangyu Chen, Jixiang Luo, Jingyu Xu, Fangqiu Yi, Chi Zhang, Xuelong Li *}
\affiliation[]{Institute of Artificial Intelligence (TeleAI), China Telecom}
\date{December 30, 2025}

\begin{document}

\abstract{
\textbf{Whether a video can be compressed at an extreme compression rate as low as  0.01\%}? To this end, we achieve the compression rate as \textbf{0.02\%} at some cases by introducing \textbf{Generative Video Compression (GVC)}, a new framework that redefines the limits of video compression by leveraging modern generative video models to achieve extreme compression rates while preserving a perception-centric, task-oriented communication paradigm, corresponding to Level C of the Shannon–Weaver model. Besides, \textbf{How we trade computation for compression rate or bandwidth}? GVC answers this question by shifting the burden from transmission to inference: it encodes video into extremely compact representations and delegates content reconstruction to the receiver, where powerful generative priors synthesize high-quality video from minimal transmitted information. \textbf{Is GVC practical and deployable}? To ensure practical deployment, we propose a compression–computation trade-off strategy, enabling fast inference on consumer-grade GPUs. Within the AI Flow framework, GVC opens new possibility for video communication in bandwidth- and resource-constrained environments such as emergency rescue, remote surveillance, and mobile edge computing. Through empirical validation, we demonstrate that GVC offers a viable path toward a new effective, efficient, scalable, and practical video communication paradigm.

\metadata[Keywords]{generative video compression, task-oriented communication, AI Flow}


\metadata[
* Correspondence to]{Xuelong Li (\email{xuelong\_li@ieee.org})\\
Other authors are listed  alphabetically by surname.
}


}

\maketitle

\section{Introduction}

\begin{itemize}
    \item \textit{\textbf{Is it possible to reconstruct high-quality video at a compression rate as low as 0.01\%}?
    \item \textbf{How we trade computation for compression rate to achieve extreme compression}?
    \item \textbf{Can extreme compression be practical and deployable in real-world scenarios}?
}
\end{itemize}

The above questions challenge the conventional paradigm of video compression. With the rapid expansion of high-resolution video, virtual reality, social media, and remote conferencing applications, video data is growing exponentially, placing unprecedented demands on existing technology and infrastructures for video storage and transmission. In bandwidth-constrained and latency-sensitive environments, achieving more efficient video compression has become a key research focus at the intersection of communications and artificial intelligence. 

Traditional communication theory, rooted in the Shannon–Weaver model introduced in the 1940s~\citep{shannon1948mathematical}, conceptualizes communication across \textbf{three levels}: \textbf{Level A} addresses the \textit{technical problem}, i.e., data-oriented communication - how to transmit information accurately; \textbf{Level B} considers the \textit{semantic problem}, i.e., semantic communication - whether the transmitted symbols convey the intended meaning; and \textbf{Level C} focuses on the \textit{effectiveness problem}, i.e., task-oriented communication - whether the received information leads to the desired behavior. 
For decades, video communication technology has primarily focused on Level A - maximizing signal fidelity under constrained bandwidth. 
Such an approach optimizes rate–distortion but can be wasteful when the receiver only needs task-relevant content rather than pixel-perfect reconstructions.

To bridge the gap between bit-level fidelity and task-level utility, the AI Flow framework \citep{aiflow} was first proposed by TeleAI at the end of 2024, which envisions leveraging communication networks to distribute intelligence for ubiquitous AI-powered services.
In addition, the Information Capacity~\citep{yuan2025information} has been proposed to evaluate the effectiveness of the generative models in data compression.
Such an advance lays the theoretical and methodological groundwork for data compression based on generative models.
In early 2025, we extended this line of work by introducing task-oriented communications for multimodal understanding via device-edge co-inference \citep{11222951_yuan_task_oriented}.
By mid-2025, TeleAI first introduced the concept of \textbf{Generative Video Compression (GVC)} in World Artificial Intelligence Conference (WAIC). 
Unlike traditional codecs that emphasize pixel-level reconstruction fidelity, GVC adopts a task-oriented communication perspective. 
It prioritizes whether the transmitted information meets perceptual expectations or effectively supports downstream tasks, thereby placing Level C at the core of its design.
At the WAIC conference, TeleAI released a prototype\footnote{https://mp.weixin.qq.com/s/QrFAmGjQvHmgEgX9En4MjQ} for maritime communications that enable ultra-low bitrate video transmission over limited-bandwidth satellite connections.
The underlying theoretical foundation has been elaborated in the technical report \citep{an2506ai}.




The core principle of GVC is \textbf{trading computation for compression rate}. Recent advances in generative models, particularly generative video models~\citep{sora,wan}, present unprecedented opportunities in video compression. By leveraging powerful generative priors, GVC aims to overcome the long-standing trade-offs between bitrate and perceptual quality found in traditional standards like HEVC~\citep{hevc}. Besides we clarify the motivation of GVC from \citep{Fan2025Infra}, which depicts the relationship among computation, bandwidth and memory, we depict the motivation of GVC. A useful metaphor illustrates this shift: traditional compression is akin to photographing a painting and sending the image; GVC, in contrast, describes the painting’s composition and style, then relies on an “AI painter” at the receiver to recreate it. Thanks to their expressive generative capabilities, modern models can synthesize high-quality videos from minimal latent representations - or even pure noise - guided by learned priors. As a result, the encoder’s role transitions from preserving every pixel to transmitting only the most task-relevant information.

As illustrated in Figure~\ref{fig:bandwidth_comparison}, GVC can achieve visually compelling reconstruction at bitrates as low as \textbf{0.005 bpp (equivalent to a 0.02\% compression rate)}. This demonstrates that GVC is advancing toward the extreme compression frontier of 0.01\%. Moreover, even when considering average performance across the common test sequences, GVC significantly reduces the required transmission bandwidth while maintaining high perceptual quality, as shown in Section~\ref{sec:results}. This makes it particularly suitable for scenarios demanding high communication efficiency, such as maritime communication, emergency rescue, narrowband mobile networks, remote video surveillance, in-vehicle or wearable devices. 

However, extreme compression introduces new challenges. High-quality reconstruction relies on computationally intensive generative processes, imposing strict requirements on hardware and inference latency. To address this, we propose the concept of \textbf{trading compression rate for practicality} - sacrificing a small fraction of the compression rate to achieve a more favorable balance between compression, computation, and quality. As shown in Table~\ref{tab:hardware_performance}, our system can run on \textbf{consumer-grade GPUs with inference latency around 2 seconds} - comparable to large language model response times - demonstrating promising real-world usability. 

This report presents a generative video compression framework tailored for ultra-low bitrate video transmission. By leveraging powerful generative video priors, the framework achieves high perceptual quality while drastically reducing bandwidth consumption with support for downstream tasks. We describe the system architecture, design strategies, and experimental validations of GVC, aiming to lay the foundation for the next generation of perception-driven video communication technology.

\section{Methodology}
\subsection{Framework Overview}
The Generative Video Compression (GVC) framework achieves high-efficiency video compression by transforming raw video frames into compact latent representations and reconstructing them through generative modeling. As illustrated in Figure~\ref{fig:framework}, the framework is composed of two primary components: a Neural Encoder and a Generative Video Decoder. 

\begin{figure}[!t]
\centering
\includegraphics[width=1\textwidth]{framework.pdf}
\caption{Overview of Our GVC Framework Grounded in the Shannon-Weaver model~\citep{shannon1948mathematical}.
Top-left: Level A addresses the technical problem, optimizing signal fidelity under limited bandwidth by minimizing distortion between input and output videos.
Top-right: Level B focus on the semantic problem, aiming at transmitting the precise semantic symbols. Bottom: Level C, central to the proposed Generative Video Compression (GVC) framework, emphasizing task-oriented effectiveness. It ensures that the compressed tokens enable the achievement of task goals - such as high-quality perception reconstruction or support for downstream tasks like segmentation.}
\label{fig:framework}
\end{figure}

The system begins by ingesting an input video sequence, which may include various types of content such as surveillance footage, video call streams, or live broadcasts. This input video is processed by the Neural Encoder, a pre-trained neural network designed to compress the video into a set of compact representations, referred to as compressed tokens.
These compressed tokens comprise both discrete and continuous representations, including compressed keyframes, high-level descriptors of video segments, and low-level continuous features. The encoder is capable of significantly reducing dimensions of the video data while preserving essential semantic information and motion dynamics. To further improve compression efficiency, the tokens are additionally encoded into a bitstream using techniques such as residual coding, reducing storage and transmission requirements. 
On the decoder side, a pre-trained diffusion-based generative video model reconstructs the video from the compressed tokens. Some of these tokens serve as direct inputs to the denoising process, while others function as the conditions. This reconstruction process is essentially a conditional video generation task, where the model synthesizes video frames that are visually faithful to the original input.
The final output is a reconstructed video that closely resembles the original in visual quality, with minimal perceptual loss, thereby achieving a balance between compression rate and visual quality.

\subsection{Trading Computation for Compression Rate}
A core idea in GVC is trading computation for compression rate. Instead of transmitting detailed visual data, GVC leverages powerful generative models at the decoder to reconstruct video content, thereby significantly reducing the bitrate required for transmission. Traditional codecs are designed to preserve signal fidelity under bitrate constraints using handcrafted signal processing techniques. In contrast, GVC shifts the burden of reconstruction to the decoder, using computation and prior knowledge embedded in generative models to synthesize realistic frames from minimal inputs.

This shift can be illustrated metaphorically: traditional compression is like photographing a painting and sending the photo; GVC is like describing the painting’s composition and style, then having an “AI painter” recreate it. Modern generative models are capable of producing high-quality video given only latent representations or even random noise, thanks to strong learned priors. Therefore, the encoder’s role becomes one of selecting and transmitting the most task-relevant information, rather than preserving every pixel.

Critically, this means that what is transmitted depends on the purpose of the reconstructed video. If the goal is human perception, the encoder transmits features that help generate perceptually similar content. If the goal is machine understanding (e.g., segmentation, recognition), then the encoder focuses on transmitting semantically meaningful representations. This represents a departure from fidelity-oriented compression toward task-oriented or effectiveness-oriented communication, aligning GVC with higher-level objectives beyond simple reconstruction accuracy.

\subsection{Trading Compression Rate for Practicality}
While trading computation for compression rate enables highly compact video representations through decoder-side generation, this approach faces practical limitations in real-world deployment. Specifically, the computational capacity of the decoder - constrained by hardware resources, power consumption, and latency requirements - imposes an upper bound on how much computation can be traded for compression. In many applications, such as real-time video conferencing or edge-device streaming, decoder-side latency and efficiency become critical bottlenecks. Consequently, the balance among reconstruction quality, compression rate, and compution must be re-evaluated with practicality as a core consideration.

To address this, our framework incorporates strategies that consciously trade compression rate for practicality, ensuring that decoding remains feasible with acceptable reconstruction quality. One such strategy is to increase the richness of the compressed latent representations, thereby reducing the reliance on large generative models at the decoder. This unlocks the ability to use smaller, faster models. Additionally, we apply model compression techniques to reduce the size and complexity of key components (e.g., 3D VAEs), and employ distillation and sampling acceleration methods for diffusion-based decoders to lower inference time. In these cases, we often compensate for the quality loss due to model simplification by transmitting higher-dimensional or more informative features, striking a new balance in the compression rate-computation-quality triangle.

Ultimately, this trade-off reflects a practical extension to the GVC paradigm: while generative models enable extreme compression, real-world usability demands adaptive strategies that scale with available computational resources. By flexibly adjusting the amount of transmitted information and the complexity of generative inference, our framework ensures that GVC remains not only efficient in terms of bitrate, but also viable and responsive under practical deployment conditions.

\section{Results}
\label{sec:results}
\begin{figure}[htbp]
    \centering
    \includegraphics[width=1.0\textwidth]{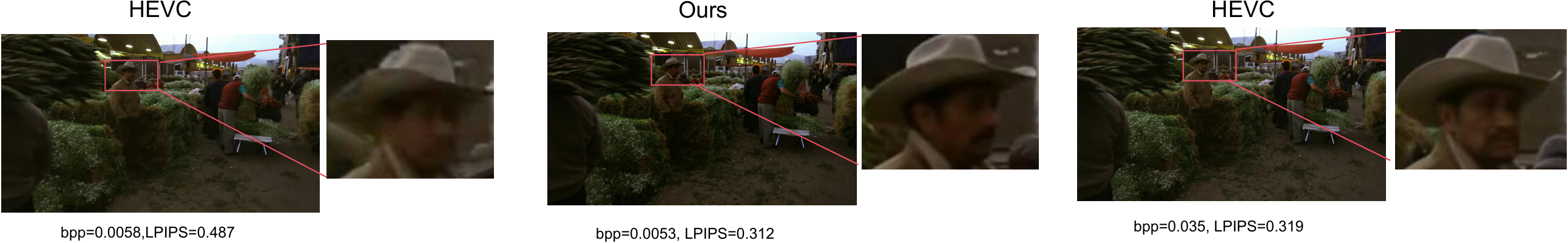}
    \caption{Bandwidth comparison for achieving comparable reconstruction quality. Traditional methods require more than 6 times the bandwidth to match the perceptual quality of our approach across selected representative sequences.}
    \label{fig:bandwidth_comparison}
\end{figure}

To validate the effectiveness of our GVC framework, we first assess the video compression performance based on a 14B video generative model on the standard benchmark: MCL-JCV~\citep{Wang_2016_MCL_JCV}. We employ mainstream perceptual metric for evaluation:\textit{ Learned Perceptual Image Patch Similarity} (LPIPS), as it is recognized as measures of human perception quality. 
As shown in Table~\ref{tab:mcljcv_results}, at an average bitrate of \textbf{0.008 bpp}\footnote{Note that the average bpp is calculated by first averaging over each sequence in the dataset and then averaging across all sequences. When testing each sequence, the trailing frames that do not form a complete GOP are discarded.}, our method maintains competitively high perceptual quality. 
In contrast, conventional video coding schemes exhibit a substantial performance gap at this bitrate. For certain challenging sequences, conventional methods need approximately \textbf{6 times} higher bitrate than our approach to attain equivalent perceptual reconstruction quality, as shown in Figure~\ref{fig:bandwidth_comparison}.

\begin{table}[ht]
\centering
\caption{Quantitative comparison on the MCL-JCV dataset. Lower values are better.}
\label{tab:mcljcv_results}
\begin{tabular}{lc}
\toprule
\textbf{Method} & \textbf{LPIPS $\downarrow$} \\
\midrule
HEVC~\cite{hevc}  & 0.278 \\
\textbf{Ours} &  0.180 \\ 
\bottomrule
\end{tabular}
\end{table}

To further validate its practical utility, we apply the reconstruction results of the model to the downstream task: video object segmentation (VOS) on DAVIS2017~\citep{Pont_Tuset_2017_DAVIS}. We evaluate performance using the Jaccard index $\mathcal{J}$, contour accuracy $\mathcal{F}$, their average ($\mathcal{J}$\&$\mathcal{F}$), and contour recall ($\mathcal{F}$-Recall). As shown in Table~\ref{tab:downstream}, our method achieves highly competitive performance. This indicates that even at low bitrates, our approach can preserve correct semantic transmission.

\begin{table*}[h]
    \centering
    \caption{Downstream performances of different coding methods. ‘\textcolor{gray}{Upper-bound}’ is obtained by evaluating the task models with the original videos. \label{tab:downstream}}
    \begin{tabular}{lcccc}
    \toprule
    \multirow{2}{*}{Method}  & \multicolumn{4}{c}{\textbf{VOS: XMEM on DAVIS2017}} \\
  \cmidrule(lr){2-5} & $\mathcal{J}\&\mathcal{F}$ (\%) & $\mathcal{J}$ (\%) & $\mathcal{F}$ (\%) & $\mathcal{F}$-Recall (\%) \\
\midrule

    HEVC@bpp=0.01  & 57.68 & 56.84 & 58.51 & 67.44 \\
    Ours@bpp=0.01 & \textbf{75.22} & \textbf{71.17} & \textbf{79.28} & \textbf{91.87} \\
    \hdashline 
    \rule{0pt}{2.5ex} \textcolor{gray}{Upper-bound}  & \textcolor{gray}{87.70} & \textcolor{gray}{84.06} & \textcolor{gray}{91.33} & \textcolor{gray}{97.02} \\
    \bottomrule
    \end{tabular}
\end{table*}

We have dedicated our effort to improving computational efficiency through techniques that include model miniaturization, knowledge distillation, and quantization. These optimizations make our approach feasible for deployment on various hardware platforms. As demonstrated in Table~\ref{tab:hardware_performance}, which reports the latency of our miniaturized model for generating a GOP of 29 frames (i.e., 29 frames at once) across different platforms, our system achieves practical inference speeds even on consumer-grade hardware.

\begin{table}[h]
\centering
\caption{Model Computational Efficiency and Hardware Performance (GOP=29)}
\label{tab:hardware_performance}
\begin{tabular}{c|l|c|c|c}
\hline
\multirow{2}{*}{\textbf{Resolution}} & \multirow{2}{*}{\textbf{Module}} & \multicolumn{3}{c}{\textbf{Latency (s)}} \\
\cline{3-5}
 & & \textbf{4090} & \textbf{A100} & \textbf{H200} \\
\hline
\multirow{2}{*}{480p} & Encoder & 0.95 & 0.64 & 0.2 \\
\cline{2-5}
 & Decoder & 1.35 & 1.4 & 1.13 \\
\hline
\multirow{2}{*}{720p} & Encoder & 1.15 & 0.80 & 0.3 \\
\cline{2-5}
 & Decoder & 6.4 & 5.5 & 2.3 \\
\hline
\multirow{2}{*}{1080p} & Encoder & 1.59  & 0.85 & 0.5 \\
\cline{2-5}
 & Decoder & 21.5 & 18 & 6.1 \\
\hline
\end{tabular}
\end{table}

\begin{figure}[htbp]
    \centering
    \includegraphics[width=0.8\textwidth]{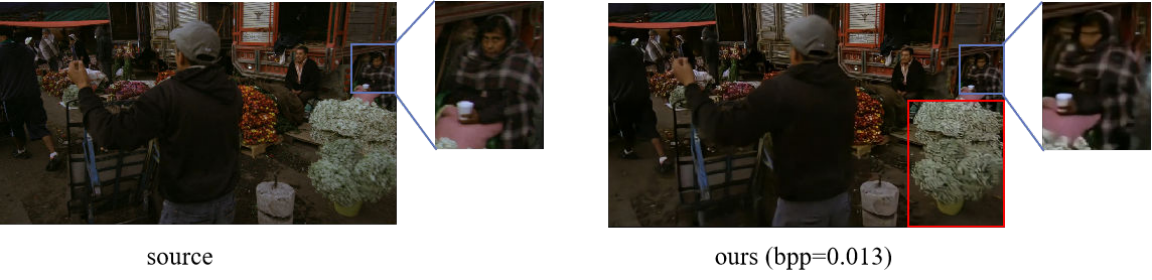}
    \caption{Visual quality comparison of the miniaturized model, demonstrating competitive perceptual quality despite model compression.}
    \label{fig:visual_quality}
\end{figure}

Although the miniaturized model incurs some loss in visual quality and bandwidth efficiency compared to its full-scale counterpart, it still maintains competitively high perceptual quality, as illustrated in Figure~\ref{fig:visual_quality}, where the shown video sequence achieves LPIPS of 
0.273. These results collectively demonstrate that our miniaturized model achieves an effective balance between computational efficiency and visual quality in practical deployment scenarios.

\section{Conclusion}
This report, under the AI Flow framework~\citep{an2506ai} and guided by the theory of Information Capacity~\citep{yuan2025information}, reimagines the foundation of video compression through the lens of Generative Video Compression (GVC) - a paradigm shift that prioritizes perceptual relevance and task effectiveness over pixel-level fidelity. By asking whether high-quality video can be reconstructed at extreme compression, we not only challenge the limits of conventional codecs but also demonstrate the feasibility of trading computation for compression in an era of increasingly capable edge devices.
Our findings show that, with the aid of modern generative video models, it is possible to achieve compelling reconstructions at extreme bitrates while maintaining visual realism and downstream task utility. 
Furthermore, we introduce the concept of trading compression rate for practicality, highlighting system designs that balance compression efficiency with inference latency and hardware constraints. Our implementation demonstrates that GVC can operate on consumer-grade GPUs with acceptable latency, making it viable for real-world deployment in domains such as remote surveillance, low-bandwidth mobile communication, and edge AI devices.
In conclusion, GVC is not just a compression technique - it embodies a task-oriented communication paradigm tailored for the era of generative intelligence. By transmitting only what is necessary for perception and decision making, it opens the door to a new class of communication systems that are more efficient, adaptive, and intelligent. We hope this work inspires further research at the intersection of generative modeling, communication theory, and real-world deployment, pushing the boundaries of what is possible in extreme video compression.

\bibliographystyle{plainnat}
\bibliography{paper}

\end{document}